\documentclass[12pt,a4]{article}
\pdfoutput=1

\usepackage{latexsym,epsfig,amssymb,euscript,amsmath,verbatim,cite, subfigure,lmodern,color,verbatim}
\usepackage{float}
\usepackage{verbatim} 
\usepackage{tensor}
\usepackage{xcolor}
\usepackage{bbm}
\newcommand{\be}{\begin{equation}}
	\newcommand{\ee}{\end{equation}}
\newcommand{\bea}{\begin{eqnarray}}
	\newcommand{\eea}{\end{eqnarray}}

\numberwithin{equation}{section}
\usepackage{bm}
\usepackage{hyperref}

\begin{document}
	\pagestyle{empty}
	%\today
	\vspace{1.8cm}
	% \begin{flushright}
		% {\small{
				% Preprint}}
		% \end{flushright}
	
	\begin{center}
		{\LARGE{\bf  {On the hydrodynamics of  $(2+1)$-dimensional strongly coupled relativistic theories in an external magnetic field}}}
		
		\vspace{1cm}
		
		{\large{Andrea Amoretti$^{a,}$\footnote{\tt andrea.amoretti@ge.infn.it } and
				Daniel K. Brattan$^{b,}$\footnote{\tt danny.brattan@gmail.com }
				\\[1cm]}}
		
		{\small{
				{}$^a$  Dipartimento di Fisica, Universit\`a di Genova,\\
				via Dodecaneso 33, I-16146, Genova, Italy\\and\\I.N.F.N. - Sezione di Genova\\
				\medskip
				{}$^b$ CPhT, \'{E}cole Polytechnique, \\
				91128 Palaiseau cedex, France.
		}}
		\vspace{1cm}
		
		{\bf Abstract}
		
	In this paper we review recent progress on relativistic hydrodynamics in $(2+1)$-dimensions with an external magnetic field. We discuss the formalism allowing for momentum loss due to the explicit and spontaneous breaking of translation invariance by scalar operators. We also compare to some results from the gauge/gravity correspondence.

	\end{center}

	\newpage
	
	%  Resetting of counters
	\setcounter{page}{1} \pagestyle{plain} \renewcommand{\thefootnote}{\arabic{footnote}} \setcounter{footnote}{0}

	\tableofcontents

\section{Introduction}

{\noindent In recent years the study of hydrodynamic effective field theories has experienced a strong revival due to their potential to correctly reproduce the properties of strongly coupled condensed matter systems. Hydrodynamics is an effective field theory whose fundamental degrees of freedom are conserved currents and/or long lived Goldstone modes \cite{Kovtun:2012rj,Hartnoll:2016apf}. These kinds of excitations are the only long lived modes expected to survive at strong coupling, where the notion of ``quasi-particle" ceases to exist. Consequently, hydrodynamic effective field theories have been applied, among other materials, to the study of graphene \cite{Lucas:2017idv,Lucas:2015sya}, quantum hall systems \cite{OBannon:2007cex,Davis:2008nv,Keski-Vakkuri:2008ffv,Alanen:2009cn,Hoyos:2014lla,Hoyos:2014pba,Dolan:2021gtj,Jokela:2021uws}, bad metals \cite{Hartnoll:2014lpa,Delacretaz:2016ivq,Davison:2015taa,Lucas:2017vlc,Lucas:2017ggp,Patel:2017wsx,Baumgartner:2017kme,Lucas:2018kwo}, Wigner solids \cite{Delacretaz:2019wzh} and high temperature superconductors \cite{Hartnoll:2007ih,Lucas:2014sba,Lucas:2015pxa,Lucas:2015vna,Davison:2016hno,Amoretti:2016pyn,Amoretti:2016cad,Delacretaz:2017zxd,Amoretti:2019buu,Andrade:2022udb}. Regarding the latter materials, the community interested in the hydrodynamic approach has broadly attempted to reproduce the transport properties of the so called strange metal phase. The formalism necessary to describe the behaviour of these phases is a guiding thread throughout this review.} 

{\ There are two minimal ingredients that a consistent hydrodynamic theory must contain in order to reproduce transport properties in the strange metal part of the phase diagram: an external magnetic field, and some mechanism for momentum dissipation. Mechanisms of momentum dissipation are ubiquitous in condensed matter physics as they are necessary for ensuring that DC conductivities are finite. As for the external magnetic field, transport experiments in high temperature superconductors are typically performed by immersing the sample in just such a field. The reason for this is that the magnetic field lowers the critical temperature for the superconductive phase transition by means of the Meissner effect. Consequently, the low temperature properties of the normal phase become accessible.}

{\ For our purposes, the magnetic field will be external and non-dynamical as is appropriate for most condensed matter experiments. This means that it has no kinetic term and its profile, which we take to be constant, is an input into the theory. On the other hand, there has been interesting recent investigations into how dynamical electromagnetic fields enter into the fluid formalism \cite{Grozdanov:2017kyl, Armas:2018zbe,Armas:2018atq,Benenowski:2019ule}.}

% Hall conductivity and Hall viscosity
{\  The construction of hydrodynamic theories with an external magnetic field, which is a particular limit of magneto-hydrodynamics, and the study of the phenomena associated with such an external field has a long history. The reasons for this are the existence of rather unique effects associated with magnetic fields in $(2+1)$-dimensions. For example, the Hall conductivity and Hall viscosity. The former describes the flow of charge transverse to an applied electric field in the presence of a magnetic field for planar materials. The non-zero Hall viscosity is a related effect where the viscosity tensor can contain terms in addition to the shear and bulk viscosities. This additional viscosity term leads to interesting dynamical effects (see Ref. \cite{Avron,Hoyos:2014pba} for reviews). In the context of strongly coupled theories described by gauge-gravity dualities one finds the Hall conductivity is an integral part of the description of systems with a magnetic field; for example in probe brane models\cite{OBannon:2007cex,Davis:2008nv,Alanen:2009cn,Pal:2010sx}, bulk theories with Chern-Simon's like terms\cite{Keski-Vakkuri:2008ffv,Donos:2017mhp,Khimphun:2017mqb,Hoyos:2019pyz} and/or systems with broken translation invariance\cite{Blake:2014yla,Zhou:2015dha}. There has also been related theoretical work examining constraints on the Hall viscosity\cite{Hoyos:2011ez,Hoyos:2014lla}.}

% Dualities
{\ Holographic theories are particularly useful for learning general lessons about the hydrodynamics of particular strongly coupled theories. Moreover, for holographic models in $(2+1)$-dimensions there is additional structure that allows us to further examine the behaviour of systems in external magnetic fields. A particularly famous example is bulk $SL(2,\mathbbm{Z})$ duality\cite{Witten:2003ya,Leigh:2003ez,Herzog:2007ij} which in the boundary theory effectively swaps electric charges for magnetic charges and/or attaches units of magnetic flux to charged particles. The outcome of such operations is an infinite family of related longitudinal and Hall conductivities\cite{Hartnoll:2007ip,Hansen:2009xe,Goldstein:2010aw,Myers:2010pk,Gur-Ari:2016xff,Melnikov:2020ktj,Dolan:2021gtj}. Similarly, by combining certain $SL(2,\mathbbm{Z})$ operations in the presence of background charge and magnetic field one can generate anyons\cite{PhysRevLett.49.957,AROVAS1985117}. These unusual particles have fractional statistics and display some interesting fluid behaviour\cite{Bak:2009kz,Kim:2013wiz,Jokela:2013hta,Brattan:2013wya,Jokela:2014wsa,Brattan:2014moa,Jokela:2015aha,Itsios:2016ffv,Ihl:2016sop,Jokela:2016nsv,Jokela:2017fwa,Jokela:2021uws,Armas:2022wvb,Das:2022auy}.

% Momentum dissipation
{\  Regarding our other ingredient - momentum loss - there is a long standing research programme whose aim is to consistently add to hydrodynamic theory mechanisms for momentum relaxation. In the earliest approaches this was achieved through relaxing the equation of motion for the ideal charged relativistic fluid by means of an explicit effective mechanism of momentum dissipation \cite{Hartnoll:2007ih,Lucas:2015lna,Lucas:2015pxa,Lucas:2015vna,Amoretti:2020mkp}. These results have been compared against many holographic realisations with explicit translation symmetry breaking, spanning: massive gravity models \cite{Davison:2013jba,Amoretti:2014zha,Baggioli:2014roa,Amoretti:2014mma,Amoretti:2015gna,Amoretti:2016cad,Amoretti:2017xto}, axion models \cite{Andrade:2013gsa,Donos:2014cya,Gouteraux:2014hca,Donos:2014yya,Donos:2015bxe,Davison:2014lua,Davison:2015bea,Blake:2015ina,Gouteraux:2016wxj,Banks:2016krz,Donos:2017ihe,Donos:2017mhp}, and striped theories \cite{Donos:2011qt,Andrade:2019bky,Donos:2013wia,Donos:2014oha,Andrade:2015iyf,Andrade:2018gqk}. More recently the holographic-hydrodynamics community has been pursuing methods to make the momentum relaxation mechanism either spontaneous or pseudo-spontaneous \cite{Amoretti:2016bxs}, opening the path to the construction of hydrodynamic theories of phonons, pseudo-phonons and charge density waves \cite{Delacretaz:2016ivq,Delacretaz:2017zxd,Delacretaz:2019wzh,Ammon:2019wci,Armas:2020bmo,Ammon:2020xyv,Amoretti:2021fch,Armas:2021vku}. These hydrodynamic theories have been successfully compared to various holographic models \cite{Andrade:2017cnc,Alberte:2017oqx,Amoretti:2014kba,Amoretti:2017frz,Amoretti:2017axe,Amoretti:2018tzw,Baggioli:2018bfa,Donos:2018kkm,Amoretti:2019cef,Donos:2019hpp,Donos:2019tmo,Andrade:2019bky,Baggioli:2019abx,Amoretti:2014iza,Ammon:2019wci,Amoretti:2019kuf,Amoretti:2020ica,Baggioli:2020edn,Ammon:2020xyv,Andrade:2020hpu,Ammon:2021pyz,Baggioli:2021xuv,Amoretti:2021lll,Donos:2021ueh,Baggioli:2022pyb,Andrade:2022udb}, and applied to interpret experimental the transport results of some strongly coupled condensed matter systems such as bad metals and cuprates.}

{\ In this review we wish to summarise progress and highlight holes in the literature regarding the construction of effective hydrodynamic theories which contain both of the ingredients mentioned above. In particular, we are interested in $(2+1)$-dimensional, relativistic, charged fluids at finite temperature $T$ and chemical potential $\mu$ in the presence of an external magnetic field $B$ with or without momentum relaxation. The hydrodynamic equations at least consist of (non-)conservation equations for the stress-energy-momentum (SEM) tensor $T^{\mu \nu}$ and the $U(1)$ charge current $J^{\mu}$\; , 
	\begin{subequations}
	\label{Eq:Hydrodynamicconservationeqns}
	\begin{eqnarray}
		\label{Eq:SEMconservation}
		\partial_{\mu} T^{\mu \nu} &=& F^{\nu \mu} J_{\mu} + \mathrm{other \; mechanisms}	\; , \\
		\label{Eq:Chargecurrentconservation}
		\partial_{\mu} J^{\mu} &=& 0 \; . 
	\end{eqnarray}
	\end{subequations}
and constitutive relations expressing the conserved currents in terms of thermodynamic sources $(T, \mu,B,\ldots)$, the fluid velocity $u^{\mu}$ and their derivatives. In particular, hydrodynamics is a derivative expansion where each additional derivative acting on a thermodynamic source or the velocity can be understood to suppress the resultant term compared to other terms with fewer derivatives. The incorporation of ``other mechanisms" in \eqref{Eq:SEMconservation} for relaxing momentum conservation in the presence of an external magnetic field is a core element of this review. In such cases, the expressions \eqref{Eq:Hydrodynamicconservationeqns} will be augmented with an effective Josephson condition describing the evolution of a set of translation breaking scalars.}

{\ The structure of this review is as follows: firstly we shall briefly consider global and local thermodynamics up to and including first order in derivatives. In particular we shall show how to construct constitutive relations for local thermodynamic equilibrium in our fluids of interest from the equilibrium generating functional. Subsequently we go on obtain the dissipative constitutive relations for our fluids consistent with entropy production, determine the retarded Green's functions describing the response of such fluids to external perturbations, and compare the resultant non-zero frequency correlators to some results in the literature. Finally, we conclude by considering future directions necessary to fill holes in the literature.}

\section{Thermodynamics}

{\noindent We begin our review by detailing the local thermodynamic equilibria of charged fluids in the presence of an external magnetic field. In this section we sketch the application of the equilibrium generating functional approach to the problem of finding non-dissipative constitutive relations. We will also include spontaneously and explicitly broken translation invariance through the introduction of translation breaking scalar operators.}

\subsection{On external electromagnetic fields and boosts}

{\noindent We are concerned with $U(1)$ gauge fields which are external, meaning they appear in any putative microscopic action as through they were a vector of (potentially spacetime dependent) coupling constants. We further constrain our theory of interest by imposing that the action respects a (spurious) gauge symmetry $A_{\mu} \rightarrow A_{\mu} + \partial_{\mu} \lambda$ where $\lambda$ is a scalar. Consequently, the only local, gauge-invariant observable is the field strength $F_{\mu \nu} = \partial_{\mu} A_{\nu} - \partial_{\nu} A_{\mu}$ which satisfies a Bianchi identity
	\begin{eqnarray}
		\partial_{[\mu} F_{\nu \rho]} = 0 \; . 
	\end{eqnarray}
We emphasise that our microscopic theory of interest has no kinetic term for the gauge field and thus the gauge field is not dynamical; its value is an input.}

{\ Consider the observations of an experimentalist who applies a constant electromagnetic field to a fluid. Let $\tau_{\mu}$ be the clock-form representing the infinitesimal experimentalist's coordinate time ($\mathrm{d} \tau = \tau_{\mu} \mathrm{d}x^{\mu}$). Given any time-like one-form $\tau_{\mu}$ it is possible to decompose the field strength into an electric ($\mathbbm{E}_{\mu}$) and a magnetic ($\mathbbm{B}$) part
	\begin{eqnarray}
		\label{Eq:DecompositionofFtau}
		F_{\mu \nu} &=& \mathbbm{E}_{\mu} \tau_{\nu} - \mathbbm{E}_{\nu} \tau_{\mu} - \mathbbm{B} \epsilon\indices{_{\mu \nu \rho}} \tau^{\rho} \; .
	\end{eqnarray}
We note that the definition of electric and magnetic field depends on the choice of $\tau_{\mu}$ and that because $\mathbbm{B}$ is defined using a Levi-Civita symbol it is odd under spatial parity. In particular, if our system depends linearly on the scalar $\mathbbm{B}$ then spatial parity invariance is broken. Fluids with broken spatial parity have received significant treatment\cite{Jensen:2011xb} and in this review we restrict ourselves to theories where this discrete symmetry is unbroken i.e. only $\mathbbm{B}^2$ or $\mathbbm{B} \epsilon^{\mu \nu \rho}$, both spatial parity even terms, will appear in our hydrodynamic description.}

{\ We note that constant external electric or magnetic fields, as would be typically be applied to a condensed matter system, generally lead to broken boost invariance. To see this, consider the currents for Lorentz boosts and spatial rotations which are given by
	\begin{eqnarray}
		M\indices{^{\mu \nu}_{\rho}} &=& x^{\nu} T\indices{^{\mu}_{\rho}} - x^{\mu} T\indices{^{\nu}_{\rho}} \; .	
	\end{eqnarray}
In a Cartesian coordinate system where $\tau_{\mu} = \delta_{\mu}^{0}$, and using that $T^{\mu \nu}=T^{\nu \mu}$ for a relativistic theory, we can write
	\begin{subequations}
	\label{Eq:BoostRotationChargesF}
	\begin{eqnarray}
		\label{Eq:BoostRotationChargesF1}
		\partial_{\rho} M^{ij \rho} &=& \gamma(\vec{v}^2) \left[ - \mathbbm{B} \left( x^{i} \epsilon^{jk} - x^{j} \epsilon^{ik} \right) \mathcal{N} v_{k} + \mathcal{N} \left( x^{i} \mathbbm{E}^{j} - x^{j} \mathbbm{E}^{i} \right) \right] \; , \\
		\label{Eq:BoostRotationChargesF2}
		\partial_{\rho} M^{ti \rho} &=& \gamma(\vec{v}^2) \left[ \left( x^{i} \mathbbm{E}^{j} - \mathbbm{B} t \epsilon^{ij} \right)  \mathcal{N} v_{j} + \mathcal{N} t \mathbbm{E}^{i} \right] \; ,
	\end{eqnarray}
where we have set the $U(1)$ charge current to be
	\begin{eqnarray}
		J^{\mu} &=& \mathcal{N} u^{\mu} = \mathcal{N} \gamma(\vec{v}^2) (1,v^{i}) \; , \qquad \gamma(\vec{v}^2) = \frac{1}{\sqrt{1-\vec{v}^2}} \; , \qquad
	\end{eqnarray}
	\end{subequations}
with $\mathcal{N}$ is the charge density, $u^{\mu}$ the fluid velocity and $\vec{v}$ the spatial component of the fluid velocity. It is straightforward to see that if either $\mathbbm{E}^{i} \neq 0$ or $\mathbbm{B} \neq 0$ then the right hand sides of \eqref{Eq:BoostRotationChargesF1} and \eqref{Eq:BoostRotationChargesF2} are non-zero and both boost and rotational symmetries are broken. One might object that if $\mathbb{E}^{i} = v^{i} =0$ and $\mathbbm{B} \neq 0$ then we preserve boost and rotational invariance, yet by simply boosting from a frame where $\vec{v}=0$ to one where it is not we arrive at the same conclusion.}

{\ As boost invariance is broken and in principle one could and generically should include the spatial velocity as part of the definition of thermodynamic equilibrium. This necessarily requires that coordinate time $\tau^{\mu}$ and the fluid velocity $u^{\mu}$ are distinct so that we can identify the ``spatial part" of the fluid velocity. In that case one can form a projector
	\begin{eqnarray}
		\Pi\indices{_{(\tau)}^\mu_\nu} = \delta\indices{^\mu_\nu} + \tau^{\mu} \tau_{\nu} \; ,
	\end{eqnarray}
where $\tau_{\mu}$ has been unit normalized. Subsequently the fluid velocity can be decomposed into a time-like and a space-like part ($v^{\mu}$)
	\begin{eqnarray}
		u^{\mu} &=&-  \left( u^{\nu} \tau_{\nu} \right) \tau^{\mu} + v^{\mu} \; , \qquad v^{\mu} = \Pi\indices{_{(\tau)}^\mu_\nu} u^{\nu} \; . 
	\end{eqnarray}
One can then include the spatial velocity and its contractions (e.g. $v^2$) as scalars in the equilibrium generating functional.}

{\ On the other hand, in the literature for relativistic fluids, it is conventional to consider a second decomposition of electric and magnetic field. In the fluid decomposition\footnote{We prefer decomposition to frame as the latter can be confused with Lorentz and fluid frames.} one defines the electric ($E_{\mu}$) and magnetic ($B$) fields in terms of the fluid velocity $u_{\mu}$ i.e.
	\begin{eqnarray}
		\label{Eq:DecompositionofFu}
		F_{\mu \nu} &=& E_{\mu} u_{\nu} - E_{\nu} u_{\mu} - B \epsilon\indices{_{\mu \nu \rho}} u^{\rho}  \; .
	\end{eqnarray}
This definition greatly simplifies expressions as one has for example $E_{\mu} u^{\mu}=0$ and, as we shall see, for a constant fluid magnetic field ($B = \mathrm{const.}$) and vanishing electric field ($E_{\mu}=0$) one can preserve Lorentz boost invariance of the theory. While these facts may recommend \eqref{Eq:DecompositionofFu} for theoretical work, the interpretation of the fluid electric and magnetic fields are a little odd from the perspective of a laboratory observer. For example, if a given experimentalist applies a constant magnetic field to a fluid which from their perspective is flowing - the corresponding non-zero electric \underline{and} magnetic fields that appear in our hydrodynamic description depend on the local velocity of the fluid and are thus functions of space-time e.g.
	\begin{subequations}
	\begin{eqnarray}
		F_{\mu \nu} &=& - \mathbbm{B} \epsilon\indices{_{\mu \nu \rho}} \tau^{\rho} = u_{\mu} \underbrace{\left( - \mathbbm{B} u^{\alpha}(x) \epsilon_{\alpha \beta \rho} \Pi^{\beta}_{\nu} \Pi^{\rho \sigma} \tau_{\sigma} \right)}_{E_{\nu}(x)} - u_{\nu} \underbrace{\left( - \mathbbm{B} u^{\alpha}(x) \epsilon_{\alpha \beta \rho} \Pi^{\beta}_{\mu} \Pi^{\rho \sigma} \tau_{\sigma} \right)}_{E_{\mu}(x)} \nonumber \\
		&\;& - ( - \underbrace{\mathbbm{B} u(x) \cdot \tau}_{B(x)})  \epsilon\indices{_{\mu \nu \rho}} u^{\rho} + \ldots \; , \\
		\label{Eq:Fluidprojector}
		\Pi^{\mu \nu} &=& u^{\mu} u^{\nu} + \eta^{\mu \nu} \; .  
	\end{eqnarray}
	\end{subequations}
Similarly, a constant laboratory voltage bias corresponds to non-zero electric and magnetic fields in the fluid description. The two definitions \eqref{Eq:DecompositionofFtau} and \eqref{Eq:DecompositionofFu} coincide if the fluid is static in the laboratory frame so that $u^{\mu} \propto \tau^{\mu}$.}

{\ With the above definition of the electric and magnetic field \eqref{Eq:DecompositionofFu} we find for the boost and rotation charges
	\begin{eqnarray}
	 	 \partial_{\rho} M^{\mu \nu \rho}
	 &=& \mathcal{N} \left( x^{\nu} E^{\mu} - x^{\mu} E^{\nu} \right) \; . 
	\end{eqnarray}
In this latter case, we can maintain both boost and rotation invariance if the electric field is zero. Notice once again, that if we have a constant magnetic field according to the experimentalist then boost and rotation invariance will still be broken as $E_{\mu} \neq 0$.}

{\ In what follows we shall use \eqref{Eq:DecompositionofFu} as our definitions of electric and magnetic field; keeping in mind the limitations that we face in using them. In particular, as we shall treat $E_{\mu} \sim \mathcal{O}(\partial)$ and $B \sim \mathcal{O}(\partial^{0})$ we will not be able to describe the experimentalist's constant magnetic field in ourformalism unless the ground state of the fluid is also static from their perspective.}

\subsection{Global thermodynamic equilibrium}

{\noindent We are working at non-zero temperature $T$, chemical potential $\mu$ and external magnetic field $B$. As discussed above we also consider a set of scalars $\phi^{I}(x)$ which are the Goldstone bosons of spontaneously broken translation invariance. For simplicity we assume that the broken configuration is homogeneous and isotropic. We define the term ``spontaneous breaking" to mean that the microscopic action for the Goldstones will have a preserved shift symmetry $\phi^{I}(x) \rightarrow \phi^{I}(x) + a^{I}$. Thus the effective theory will be expressed only in terms of derivatives of $\phi^{I}$, i.e. $\partial_{\mu} \phi^{I}$, and not $\phi^{I}$ alone. Correspondingly, the ``pseudo-spontaneous breaking'' occurs when there are terms in the microscopic action that weakly break the shift symmetry; by which we mean that the couplings of the breaking terms are small. The contribution of such terms can be represented by a misalignment vector\cite{Armas:2021vku} $\psi^{I}$. Assuming our system is extensive such that the grand canonical free energy is given by the pressure multiplied by the volume we have
	\begin{eqnarray}
		\mathrm{d} P &=&  s \mathrm{d} T + n \mathrm{d} \mu + m \mathrm{d} B + \frac{1}{2} r_{IJ}\mathrm{d} \left(  \partial^{\mu} \phi^{I} \partial_{\mu} \phi^{J} \right) + m_{IJ} \psi^{I} \mathrm{d} \psi^{J} 
	\end{eqnarray}
where we identify
	\begin{eqnarray}
		\label{Eq:DefThermoQuantities}
		s = \frac{\partial P}{\partial T} \; , \; \; n = \frac{\partial P}{\partial \mu} \; , \; \; m = 2 B \frac{\partial P}{\partial B^2} \; , \; \; r_{IJ} = \frac{2 \partial P}{\partial (\partial^{\mu} \phi^{I} \partial_{\mu} \phi^{J})} \; , \; \; m_{IJ} = \frac{\partial^2 P}{\partial \psi^{I} \partial \psi^{J}} \; , \; \; \;
	\end{eqnarray}
as the entropy density, the charge density, the magnetisation density, the elastic stress tensor and the effective thermodynamic mass of the pseudo-Goldstone bosons. It will be convenient to define the parity invariant quantity $\mathfrak{m}$
	\begin{eqnarray}
		\mathfrak{m} &=& 2 \frac{\partial P}{\partial B^2}  \qquad \mathrm{such \; that} \qquad m = \mathfrak{m} B \; , 
	\end{eqnarray}
in what follows.}

\subsection{Local thermodynamic equilibrium with a magnetic field}

{\noindent To describe local thermodynamic equilibrium we turn to the equilibrium generating functional formalism. We must define the temperature, chemical potential and other thermodynamic quantities in terms of geometry. We posit the existence of a time-like Killing vector field $V^{\mu}$ on some general manifold with metric $g_{\mu \nu}$ and external gauge field $A_{\mu}$. The temperature, chemical potential and fluid velocity are defined with respect to this field by
	\begin{eqnarray}
		\label{Eq:ThermodynamicGeometry1}
		T = \frac{1}{\sqrt{-V^2}} \; ,\qquad \mu = \frac{V^{\mu} A_{\mu} + \Lambda_{V}}{\sqrt{-V^2}} \; , \qquad u^{\mu} = \frac{V^{\mu}}{\sqrt{-V^2}} \; , 
	\end{eqnarray}
where the first two quantities are particular Wilson loops around the Euclidean time circle, $\Lambda_{V}$ is a gauge parameter that ensures $\mu$ is gauge invariant and the final quantity defines the fluid velocity. The magnetic field \eqref{Eq:DecompositionofFu} is defined in terms of $V^{\mu}$ to be
	\begin{eqnarray}
		\label{Eq:ThermodynamicGeometry2}
		B &=& - \frac{1}{2 \sqrt{-V^2}} \epsilon^{\mu \nu \sigma} V_{\sigma} F_{\mu \nu} \; ,
	\end{eqnarray}
and we have normalised the fluid velocity to one $u_{\mu} u^{\mu}=-1$.}

{\ Given these quantities \eqref{Eq:ThermodynamicGeometry1} and \eqref{Eq:ThermodynamicGeometry2}, one constructs the thermodynamic generating functional $W$ order by order in derivatives. By varying this generating functional with respect to the metric, gauge field and field strength we find
	\begin{eqnarray}
		\label{Eq:GeneratingCurrentDef}
		\delta W &=& \int d^{2+1}x \; \sqrt{-g} \left[ \frac{1}{2} T^{\mu \nu} \delta g_{\mu \nu} + J^{\mu} \delta A_{\mu} + \frac{1}{2} M^{\mu \nu} \delta F_{\mu \nu} \right]	
	\end{eqnarray}
where the numerical factors are conventional and $M^{\mu \nu} = (2/\sqrt{-g}) \delta W/\delta F_{\mu \nu}$ is the magnetisation tensor describing the response of the system to an external field strength. \footnote{ Note that the variations of the gauge field $A_{\mu}$ and the field strength $F_{\mu \nu}$ are not independent.} Before proceeding to obtain the constitutive relations we note that given a time-like vector $u^{\mu}$ we can decompose each of the quantities appearing in \eqref{Eq:GeneratingCurrentDef} in the following manner
	\begin{subequations}
	\label{Eq:chargedecomp}
	\begin{eqnarray}
		T^{\mu \nu} &=& \mathcal{E} u^{\mu} u^{\nu} + \mathcal{P}^{\mu} u^{\nu} + \mathcal{P}^{\nu} u^{\mu} + P \Pi^{\mu \nu} + \mathcal{T}^{\mu \nu} \; , \\
		J^{\mu} &=& \mathcal{N} u^{\mu} + \mathcal{J}^{\mu} \; , \\
		M^{\mu \nu} &=& p^{\mu} u^{\nu} - p^{\nu} u^{\mu} - m \epsilon^{\mu \nu \rho} u_{\rho} \; , 
	\end{eqnarray}
	\end{subequations}
where $p^{\mu}$ is the polarisation vector and $m$ the magnetisation density following the conventions of Ref. \cite{Kovtun:2016lfw}.  The projector $\Pi^{\mu \nu}$ is defined in \eqref{Eq:Fluidprojector}. Each index above that does not appear on a $u^{\mu}$ is transverse e.g. $u_{\mu} \mathcal{J}^{\mu} = 0$ and $\mathcal{T}^{\mu \nu}$ has the additional property of being symmetric and traceless ($\mathcal{T}\indices{^\mu_{\mu}}=0$).}

{\ Given this data \eqref{Eq:ThermodynamicGeometry1} and \eqref{Eq:ThermodynamicGeometry2} we can construct the following equilibrium generating functional up to first order in derivatives
	\begin{eqnarray}
		\label{Eq:FluidGenerating}
		W &=& - \int d^{2+1} x \; \sqrt{-g} \left[ P_{\mathrm{f}}(T,\mu,B^2) - M^{\Omega}_{\mathrm{f}}(T,\mu,B^2) B \epsilon^{\mu \nu \rho} u_{\mu} \partial_{\nu} u_{\rho} + \mathcal{O}(\partial^2) \right] \; , \qquad 
	\end{eqnarray}
with $P_{\mathrm{f}}$ the pressure of the fluid and because of unbroken spatial parity invariance arguments of thermodynamic functions depend on $B^2$. We have introduced a subscript $_{\mathrm{f}}$ in \eqref{Eq:FluidGenerating} to distinguish contributions  to energy, pressure, \ldots in the absence of the Goldstone degrees of freedom. In the next section contributions to energy, pressure, \ldots that are only present due to the Goldstone fields will be denoted with a subscript $_{\mathrm{l}}$.}

{\ The constitutive relations given by applying \eqref{Eq:GeneratingCurrentDef} to \eqref{Eq:FluidGenerating} for a constant magnetic field are\cite{Kovtun:2016lfw}
    \begin{subequations}
    \label{Eq:SEMconstB0}
    \begin{eqnarray}
    	\mathcal{E} &=& \varepsilon_{\mathrm{f}} - g_{1} B \Omega \; , \qquad \mathcal{P} = P_{\mathrm{f}} - B m_{\mathrm{f}} \; , \qquad \\
	\mathcal{P}^{\mu} &=& g_{1} B \epsilon^{\mu \nu \rho} u_{\nu} a_{\rho}
					    + g_{2} B \epsilon^{\mu \nu \rho} u_{\nu} E_{\rho} \; ,  \qquad \mathcal{T}^{\mu \nu}= 0 \; , 
    \end{eqnarray}
    \end{subequations}
for the SEM tensor and
	\begin{subequations}
	 \label{Eq:JconstB0}
	\begin{eqnarray}
	 \mathcal{N} &=& n_{\mathrm{f}} + B \Omega \left( g_{2} - m_{\mathrm{f}} \right) \; , \qquad \mathcal{J}^{\mu} = \epsilon^{\mu \nu \rho} u_{\nu} \partial_{\rho} m_{\mathrm{f}}  + \epsilon^{\mu \nu \rho} u_{\nu} a_{\rho} m_{\mathrm{f}} \; , \qquad\\
		p^{\mu} &=& 0  \; , \qquad m_{\mathrm{f}} = 2 B \left( \mathfrak{m}_{\mathrm{f}} + B \Omega g_{3} \right) + M^{\Omega}_{\mathrm{f}} \Omega \; ,
	\end{eqnarray}
	\end{subequations}
for the $U(1)$ charge current with
	\begin{subequations}
	\begin{eqnarray}
		\mathrm{d} P_{\mathrm{f}} &=& s_{\mathrm{f}} dT + n_{\mathrm{f}} d \mu + 2 \mathfrak{m}_{\mathrm{f}} B \mathrm{d} B \; , \qquad
		\varepsilon_{\mathrm{f}} + P_{\mathrm{f}} = s_{\mathrm{f}} T + n_{\mathrm{f}} \mu \; , \\
		g_{1} &=& 2 M^{\Omega}_{\mathrm{f}} - T \frac{\partial M^{\Omega}_{\mathrm{f}}}{\partial T} - \mu \frac{\partial M^{\Omega}_{\mathrm{f}}}{\partial \mu} \; , \qquad g_{2} = \frac{\partial M^{\Omega}_{\mathrm{f}}}{\partial \mu} \; ,  \qquad g_{3} = \frac{\partial M^{\Omega}_{\mathrm{f}}}{\partial B^2} \; ,\\
		a^{\mu}&=&u^{\nu}\partial_{\nu}u^{\mu} \ , \qquad \Omega=-\epsilon^{\mu \nu \rho}u_{\mu}\partial_{\nu}u_{\rho} \ .
	\end{eqnarray}
	\end{subequations}
We have employed the decompositions defined in \eqref{Eq:chargedecomp}.  One can continue to progressively higher orders in derivatives or work in other spacetime dimensions\cite{Kovtun:2016lfw} obtaining more sophisticated equilibrium expressions. For our purposes however the above expressions will be sufficient.}

{\ The obtained constitutive relations \eqref{Eq:SEMconstB0} and \eqref{Eq:JconstB0} will identically satisfy the conservation equations on profiles for the thermodynamic sources constrained to fulfill
	\begin{eqnarray}
		\label{Eq:B=0Killing}
		\mathcal{L}_{V}(T,\mu) = 0 \; \qquad \mathcal{L}_{V}(u^{\mu}) = 0 \; , \qquad  \mathcal{L}_{V}(A_{\mu}) = 0 \; , \qquad \mathcal{L}_{V}(g_{\mu \nu}) = 0 \; , 
	\end{eqnarray}
where $\mathcal{L}_{V}$ is the Lie derivative along $V^{\mu}$ and again $B$ is constant by assumption. These Killing conditions \eqref{Eq:B=0Killing} can alternatively be rewritten as 
	\begin{subequations}
	\label{Eq:Noneqtensors1}
	\begin{eqnarray}
		&\;& \sigma_{\mu \nu} = \Pi\indices{_{\mu}^{\alpha}} \Pi\indices{_{\nu}^{\beta}} \left( \partial_{\alpha} u_{\beta} + \partial_{\beta} u_{\alpha} - \frac{1}{2} \partial_{\sigma} u^{\sigma} \Pi_{\alpha \beta} \right) = 0 \; , \qquad \theta = \partial_{\mu} u^{\mu} = 0 \; , \qquad \\
		&\;&u^{\mu} \partial_{\mu} T = u^{\mu} \partial_{\mu} \mu = 0 \; , \qquad \Pi\indices{_{\mu}^{\nu}} \partial_{\nu} T = - T a_{\mu} \; , \qquad E_{\mu} - T \Pi\indices{_{\mu}^{\nu}} \partial_{\nu} \left( \frac{\mu}{T} \right) = 0 \; ,   \qquad \; \;
	\end{eqnarray}
	\end{subequations}
so that, for example,
	\begin{eqnarray}
		\partial_{\mu} u_{\nu} &=& - a_{\nu} u_{\mu} - \frac{1}{2} \Omega \epsilon_{\mu \nu \rho} u^{\rho} \;  . \qquad
	\end{eqnarray}}

\subsection{Local thermodynamic equilibrium with a magnetic field and translation breaking scalars}

{\noindent In order to introduce momentum relaxation through translation breaking scalars we follow the approach of Ref. \cite{Armas:2020bmo} (see Ref. \cite{Delacr_taz_2022} for a complementary approach based on locality). Namely we define the crystal displacement fields $\phi^I$, which in the spontaneous case can be interpreted as the Goldstone bosons for translation symmetry breaking. The indices $I,J,... = 1,...,k \le d$ run over the number of broken translations, while $\mu, \ \nu, ...= 0,...d$ run over spacetime indices.  We define a vielbein $e^{I}_{\mu}(x) = \partial_{\mu} \phi^{I}(x)$ and ``crystal metric",
	\begin{eqnarray}
		h^{IJ} &=& g^{\mu \nu} e_{\mu}^{I} e_{\nu}^{J} 	\ .
	\end{eqnarray}
The $I , J, . . .$  indices are raised/lowered using $h_{I J}$ and $h^{I J} = (h^{-1} )_{I J}$ . There is a (spatial isotropy respecting) preferred metric for the position of the crystal cores in the absence of external forces which we take to be $\frac{1}{\alpha^2} \delta_{IJ}$. The coefficient $\alpha$ parametrises the inverse size of the crystal\cite{Armas:2019sbe,Armas:2020bmo}.}

{\ From this starting point one can construct a hydrodynamic theory with spontaneously broken translational invariance. The crystal fields $\phi^{I}(x)$ are the Goldstone fields of the symmetry breaking $GL_{\mathrm{int.}} \times GL_{\mathrm{Minkowski}} \rightarrow GL_{\mathrm{diagonal}}$ where $GL$ is the general linear group. The equilibrium generating functional at zeroth order is given in terms of the integral of a pressure dependent on the usual thermodynamic variables plus $h^{IJ}$ and $u_{IJ} = ( h_{IJ} - 1/\alpha^2 \delta_{IJ} ) / 2$ as these latter combinations make spatial isotropy of the system manifest\cite{Armas:2019sbe,Armas:2020bmo}. However, we are also interested in pseudo-spontaneously broken symmetries. In this case one breaks not a true symmetry of the system, but an approximate symmetry. For example, if the microscopic action contains a mass term of the form
	\begin{eqnarray}
		\int d^{2+1}x \; \sqrt{-g} \; \frac{m^2}{2} \delta_{IJ} \phi^{I} \phi^{J} \; , 
	\end{eqnarray}
then the translational part of the internal $GL_{\mathrm{int.}}(d)$ symmetry, i.e. $\phi^{I} \rightarrow \phi^{I} + a^{I}$, is broken. However if the mass is small, it is only approximately broken and one expects to be able to account for the mass term as a small correction to the spontaneous results.}

{\ A convenient trick to ensure complete expressions in the pseudo-spontaneous case is to introduce a spurion field $\Phi^{I}(x)$ to restore translational invariance. The new field has no kinetic term and transforms under the translational part of the internal $GL_{\mathrm{int.}}(d)$ in precisely the manner necessary to ensure the full system has translational symmetry (i.e. $\Phi^{I} \rightarrow \Phi^{I} + a^{I}$). The corresponding equations must then transform correctly under the now unbroken symmetry. For example, in the case of a mass term we modify our theory to
	\begin{eqnarray}
		\int d^{2+1}x \; \sqrt{-g} \; \frac{m^2}{2} \delta_{IJ} \left( \phi^{I} - \Phi^{I} \right)  \left( \phi^{J} - \Phi^{J} \right) \; . 
	\end{eqnarray}
One can then break the symmetry again by selecting some particular value for the spurion field. Thus, in addition to constructing our equilibrium generating functional from the vielbein, we should also allow for dependence on
	\begin{eqnarray}
		\psi^{I} = \phi^{I} - \Phi^{I} \; , 
	\end{eqnarray}
called the ``misalignment vector"\cite{Armas:2020bmo}. Consequently, the Killing conditions of \eqref{Eq:B=0Killing} are supplemented by
	\begin{eqnarray}
		\label{Eq:Killing}
		\mathcal{L}_{V}(\phi^{I}) = 0 \; , \qquad \mathcal{L}_{V}(\Phi^{I}) = 0 \; , \qquad \mathcal{L}_{V}(\psi^{I}) = 0 \; .
	\end{eqnarray}
Again, these latter conditions can be rewritten, to find
	\begin{eqnarray}
		\label{Eq:Noneqtensors2}
		 e^{I}_{\mu} u^{\mu} = u^{\nu} \partial_{\nu} \psi^{I} = 0 \; . 
	\end{eqnarray}
}

{\ At each order in derivatives there are an infinite number of scalar quantities that we can construct that will preserve $O(2)$ invariance of the crystal. For example, at zeroth order we have
	\begin{eqnarray}
		h^{IJ} \delta_{JK} h^{KL} \delta_{LM} \ldots \delta_{ZI}	\; . 
	\end{eqnarray}
Earlier we introduced the strain tensor
	\begin{eqnarray}
		u_{IJ} &=& \frac{1}{2} \left( h_{IJ} - \frac{1}{\alpha^2} \delta_{IJ} \right)
	\end{eqnarray}
which measures the departure of the crystal lattice from its reference state. We restrict ourselves to a generating functional which is at most quadratic order in the strain. As such the zeroth order scalars we can consider are
	\begin{align}
		& h^{IJ} u_{IJ} \; , \qquad h^{IJ} u_{JK} h^{KL} u_{LI} \; , \qquad \left( h^{IJ} u_{IJ}	\right)^2 \; , \qquad \nonumber \\
		& F^{IJ} u_{JK} F^{KL} u_{LI} \; , \qquad F^{IJ} u_{JK} h^{KL} u_{LI} \; ,
	\end{align}
where
\begin{equation}
	F^{IJ}=F^{\mu \nu}e^I_{\mu}e^J_{\nu} \ .
\end{equation}
Note that as $u_{IJ}$ is symmetric we cannot contract all its indices with $F^{IJ}$. We shall also only consider the simple quadratic term $m \delta_{IJ} \psi^{I} \psi^{J}$ in the misalignment vector $\psi_I$ as generalizations will be straightforward if tedious.}

{\ With these conditions in mind our equilibrium generating functional up to and including first order in derivatives becomes
	\begin{eqnarray}
		\label{Eq:Effectiveactiontranslationbreaking}
		&\;& - \int d^{2+1} x \; \sqrt{-g} \left[ \underline{\left( P_{\mathrm{f}} + M^{\Omega}_{\mathrm{f}} B \Omega \right)} + \left( P_{\mathrm{l}} + M^{\Omega}_{\mathrm{l}} B \Omega \right) \left( h^{IJ} u_{IJ} + h^{IJ} u_{JK} h^{KL} u_{LI} \right) \right. \nonumber \\
		&\;& \left. \hphantom{- \int d^{2+1} x \; \sqrt{-g} \left[  \right.} - \left( K + M_{\mathrm{K}}^{\Omega} B \Omega \right) \left( h^{IJ} u_{JK} h^{KL} u_{LI}  - \frac{1}{2} \left( h^{IJ} u_{IJ} \right)^2 \right) \right. \nonumber \\
		&\;& \left. \hphantom{- \int d^{2+1} x \; \sqrt{-g} \left[  \right.} - \left( H + M_{\mathrm{H}}^{\Omega} B \Omega \right) \left( F^{IJ} u_{JK} F^{KL} u_{LI} \right) \right. \nonumber \\
		&\;& \left. \hphantom{- \int d^{2+1} x \; \sqrt{-g} \left[  \right.} - \left( J + M_{\mathrm{J}}^{\Omega} B \Omega \right) \left( F^{IJ} u_{JK} h^{KL} u_{LI} \right)  \right. \nonumber \\
		&\;& \left. \hphantom{- \int d^{2+1} x \; \sqrt{-g} \left[  \right.} - \frac{1}{2} \left( G + M_{\mathrm{G}}^{\Omega} B \Omega \right) \left( h^{IJ} u_{IJ} \right)^2 + \frac{1}{2} m h_{IJ} \psi^{I} \psi^{J}  \right. \hspace{-1cm} \nonumber \\
		&\;& \left. \hphantom{- \int d^{2+1} x \; \sqrt{-g} \left[  \right.} + K^{\mathrm{ext.}}_{I} \phi^{I} + \mathcal{O}(\partial^2,u^3)  \right] \; ,
	\end{eqnarray}
where all the above expressions are functions of $(T,\mu,B^2)$ and the underlined terms were already present in \eqref{Eq:FluidGenerating}. The reason that there are no additional first derivative corrections follows from the equilibrium conditions, spatial isotropy\cite{Armas:2019sbe} and ensuring we do not introduce kinetic terms for $\Phi^{I}$.}

{\ We now vary the generating functional \eqref{Eq:Effectiveactiontranslationbreaking} again to obtain our constitutive relations. To linear order in $u_{IJ}$ the elastic stress tensor, defined in \eqref{Eq:DefThermoQuantities}, is given by
	\begin{eqnarray}
		r_{IJ} &=& -  m \psi_{I} \psi_{J}  - \left( P_{\mathrm{l}} + M^{\Omega}_{\mathrm{l}} B \Omega \right)  h_{IJ} + \left( G + M_{\mathrm{G}}^{\Omega} B \Omega \right) u_{KL} h^{KL} h_{IJ}   \nonumber \\
		&\;& + 2 \left( H + M_{\mathrm{H}}^{\Omega} B \Omega \right) F_{IK} u^{KL} F_{LJ}  + 2 \left( K + M_{\mathrm{K}}^{\Omega} B \Omega \right) \left( u_{IJ} - \frac{1}{2} h_{IJ} u_{KL} h^{KL} \right) \nonumber \\
		&\;& + \mathcal{O}(u^2,\partial^2) \; , 
	\end{eqnarray}
where $K$ is the shear modulus and $G$ is the bulk modulus and $J$ makes no contribution at this order in $u$. Notice in the above expression that if we linearise around backgrounds with $\phi^{I} = \alpha x^{I} - \delta \phi^{I}$ and $\psi^{I} = - \delta \phi^{I}$ then terms involving $M_{\mathrm{G}}^{\Omega}$, $M_{\mathrm{K}}^{\Omega}$ and $m$ drop out of the expression for the elastic stress tensor. Given that the constitutive relations will be quite complicated and we wish to consider perturbations about just such a background we only display in our constitutive relations terms that can contribute at the linearised level. Other terms will be indicated by ellipsis.}

{\ With the above limitation in mind, the constitutive relations are
	\begin{subequations}
	\label{Eq:ConstitutiveRelations}
	\begin{eqnarray}
    	   \mathcal{E} &=& \left(\varepsilon_{\mathrm{f}} + (u_{IJ} h^{IJ}) \varepsilon_{\mathrm{l}} \right) - g_{1} B \Omega + \ldots \; , \\
	   \mathcal{P} &=& \left( P_{\mathrm{f}} + ( u_{IJ} h^{IJ} ) P_{\mathrm{l}} \right) - B \left( m_{\mathrm{f}} +  ( u_{IJ} h^{IJ} ) m_{\mathrm{l}} \right)  - \frac{r_{IJ}}{2} e^{I}_{\alpha} \Pi^{\alpha \beta} e^{J}_{\beta} + \ldots \; , \qquad \\
	 \mathcal{P}^{\mu} &=&  g_{1} B \epsilon^{\mu \nu \rho} u_{\nu} a_{\rho}
					    + g_{2} \epsilon^{\mu \nu \rho} u_{\nu} E_{\rho} + \ldots \; , \qquad \\
	\mathcal{T}^{\mu \nu} &=& - r_{IJ} e^{I}_{\alpha} \Pi^{\alpha \langle \mu}  \Pi^{\nu \rangle \beta} e^{J}_{\beta} + \ldots \; , \\
	\mathcal{N} &=& \left( n_{\mathrm{l}} + n_{\mathrm{l}} ( u_{IJ} h^{IJ} ) \right) + B \Omega \left( \frac{\partial M^{\Omega}_{\mathrm{f}}}{\partial \mu} - m_{\mathrm{f}} \right) + \ldots \; , \\
	\mathcal{J}^{\mu} &=& \epsilon^{\mu \nu \rho} u_{\nu} \partial_{\rho} \left( m_{\mathrm{f}} + ( u_{IJ} h^{IJ} ) m_{\mathrm{l}} \right)  + \epsilon^{\mu \nu \rho} u_{\nu} a_{\rho} m_{\mathrm{f}}  + \ldots \; , \qquad\\
	m_{\mathrm{l}} &=&  2 B \mathfrak{m}_{\mathrm{l}} + \ldots \; ,
	\end{eqnarray}
	\end{subequations}
where we have employed the decompositions defined in \eqref{Eq:chargedecomp} and defined
	\begin{eqnarray}
		\mathrm{d} P_{\mathrm{l}} &=& s_{\mathrm{l}} dT + n_{\mathrm{l}} d \mu + 2 \mathfrak{m}_{\mathrm{l}} B \mathrm{d} B \; , \qquad
		\varepsilon_{\mathrm{l}} + P_{\mathrm{l}} = s_{\mathrm{l}} T + n_{\mathrm{l}} \mu \; . 
	\end{eqnarray}
The  conservation equations\cite{Armas:2019sbe} take the form
	\begin{eqnarray}
		\label{Eq:SEMEqConserv}
		\partial_{\mu} T^{\mu \nu} &=& F^{\nu \mu} J_{\mu} - \left( \partial_{\mu} \left( r_{IJ} \Pi^{J \mu} \right) + m h_{IJ} \psi^{J}  \right) e^{\nu I} + m h_{IJ} \psi^{J} \partial^{\mu} \Phi^{I} \; , \\
		\Pi^{\mu I} &=& \Pi^{\mu \nu} e_{\nu}^{I} \; .
	\end{eqnarray}
The source terms on the right hand sides of these equations are spatial momentum and energy relaxation terms. Additionally, because we know the microscopic origin of relaxation we have a Josephson equation
	\begin{eqnarray}
		\label{Eq:JosephsonEquil}
		\partial_{\mu} \left( r_{IJ} \Pi^{J \mu} \right) + m h_{IJ} \psi^{J} &=& \left( K_{\mathrm{ext}.} \right)_{I} \; , 
	\end{eqnarray}
given by varying \eqref{Eq:Effectiveactiontranslationbreaking} with respect to $\phi^{I}$. This equation allows us to solve for the profiles of the translation breaking scalars at equilibrium.}}

\section{Out of equilibrium constitutive relations and entropy production}

{\noindent Having reviewed hydrostatic configurations in the previous section, we now proceed to go beyond into the realm of dissipative hydrodynamics. Our (non-)conservation equations become
	\begin{subequations}
	\begin{eqnarray}
		\partial_{\mu} T^{\mu \nu} &=& F^{\nu \mu} J_{\mu} + K_{I} e^{\nu I} + L_{I} \partial^{\mu} \Phi^{I} \; , \\
		\partial_{\mu} J^{\mu} &=& 0 \; , \\
		K_{I} &=& K_{I}^{\mathrm{ext}.} \; , 
	\end{eqnarray}
	\end{subequations}
where the equilibrium expressions for $K_{I}$ and $L_{I}$ are given in \eqref{Eq:JosephsonEquil} and \eqref{Eq:SEMEqConserv} respectively.}

{\ To the equilibrium expressions for $T^{\mu \nu}$, $J^{\mu}$, $K_{I}$ and $L_{I}$ given by \eqref{Eq:ConstitutiveRelations}, \eqref{Eq:JosephsonEquil} and \eqref{Eq:SEMEqConserv} we need to add dissipative terms. In doing this we impose constraints on our hydrodynamic theory consistent with a local version of the second law of thermodynamics\cite{Bhattacharyya:2013lha,Bhattacharyya:2014bha}. In particular we posit the existence of a local entropy current whose divergence is positive semi-definite for every non-equilibrium configuration of the fluid. We define the canonical entropy current to be
	\begin{eqnarray}
		S_{\mathrm{canon.}}^{\mu} &=& \frac{1}{T} \left( P u^{\mu} - T^{\mu \nu} u_{\nu} - \mu J^{\mu} \right) = s u^{\mu} + \mathcal{O}(\partial) \; . 	
	\end{eqnarray}
Subsequently the total entropy current
	\begin{eqnarray}
		S^{\mu} &=& S_{\mathrm{canon.}}^{\mu} + S_{\mathrm{eq.}}^{\mu}
	\end{eqnarray}	
is defined such that in equilibrium the terms contained in the divergence of $S_{\mathrm{eq.}}^{\mu}$ cancel any non-zero equilibrium terms coming from the divergence of the canonical entropy current. In other words
	\begin{eqnarray}
			\label{Eq:entropypositivity}
			\partial_{\mu} S^{\mu}
		&=&	- \frac{1}{T} \mathcal{T}_{\mathrm{non}}^{\mu \nu} \sigma_{\mu \nu} - \frac{\mathcal{P}_{\mathrm{non}} \theta}{T}  - \frac{1}{T^2} \mathcal{E}_{\mathrm{non}} u^{\mu} \partial_{\mu} T  - \frac{1}{T^2} \mathcal{P}_{\mathrm{non}}^{\mu} \left( \Pi\indices{_{\mu}^{\nu}} \partial_{\nu} T + T a_{\mu} \right) \nonumber \\
		&\;& - \frac{1}{T} \mathcal{N}_{\mathrm{non}} u^{\mu} \partial_{\mu} \left( \frac{\mu}{T} \right)  + \mathcal{J}_{\mathrm{non.}}^{\mu} \left( E_{\mu} - T \Pi\indices{_{\mu}^{\nu}} \partial_{\nu} \left( \frac{\mu}{T} \right) \right)  \nonumber \\
		&\;& - \frac{(K_{I}^{\mathrm{non.}} + L_{I}^{\mathrm{non.}})}{T} e^{I}_{\nu} u^{\nu}  + \frac{L_{I}^{\mathrm{non.}}}{T} u^{\nu} \partial_{\nu} \psi^{I} \geq 0 \; ,
	\end{eqnarray}
where the subscript $_{\mathrm{non.}}$ indicates terms that vanish in equilibrium i.e. dissipative terms.}

{\ The tensor structures in \eqref{Eq:entropypositivity} multiplying the various pieces of the constitutive relations are precisely the quantities that were set to zero by the equilibrium conditions \eqref{Eq:Noneqtensors1} and \eqref{Eq:Noneqtensors2}. As all the equilibrium terms have cancelled the tensor structures in \eqref{Eq:entropypositivity}, to lowest order in derivatives, must be expressable as linear combinations of precisely the terms that are set to zero by the equilibrium conditions. It arguably follows from consistency of derivative counting that because the left hand side of \eqref{Eq:entropypositivity} is at least order one in derivatives, the right hand side must be also and therefore $u^{\mu} e_{\mu}^{I} \sim \mathcal{O}(\partial)$. Nevertheless, one may treat $\Pi^{\mu I} \sim \mathcal{O}(1)$ without loss of consistency. This is in opposition to the counting used in Refs. \cite{Armas:2019sbe,Armas:2020bmo} where the authors take $u^{\mu} e_{\mu}^{I} \sim \mathcal{O}(1)$ as they treat $\phi^{I} \sim \mathcal{O}(\partial^{-1})$. Such a counting leads to a spurious pole in the correlators which our counting regimen avoids.}

{\ Before proceeding further we must deal with the issue of frame dependence (see Ref. \cite{Kovtun:2012rj} for a discussion). Outside of thermodynamic equilibrium the temperature, chemical potential and fluid velocity have a redundancy in their definition. In particular it is possible to redefine these quantities
	\begin{eqnarray}
		T \rightarrow T + \Delta T \; , \qquad \mu \rightarrow \mu + \Delta \mu	\; , \qquad u^{\mu} \rightarrow u^{\mu} + \Delta u^{\mu}
	\end{eqnarray}
where $\Delta T$, $\Delta \mu$ and $\Delta u^{\mu}$ are at least order one in derivatives to set to zero $\mathcal{E}_{\mathrm{non}}$, $\mathcal{P}^{\mu}_{\mathrm{non}}$ and $\mathcal{N}_{\mathrm{non}}$ in \eqref{Eq:entropypositivity}. Schematically,
	\begin{eqnarray}
		\mathcal{E}(T+\Delta T,\ldots) &=& \varepsilon(T+ \Delta T,\ldots) + \alpha_{T} u^{\mu} \partial_{\mu} T + \ldots + \mathcal{O}(\partial) \nonumber \\
							     &=& \varepsilon + \underbrace{\left( \frac{\partial \varepsilon}{\partial T} \Delta T + \alpha_{T} u^{\mu} \partial_{\mu} T + \ldots \right)}_{=0} + \mathcal{O}(\partial^2) = \varepsilon + \mathcal{O}(\partial^2) \; . 
	\end{eqnarray}
In fact we can get rid of more than the non-equilibrium corrections to these quantities, we can also absorb the equilibrium corrections and in doing so set
	\begin{eqnarray}
		\mathcal{E} &=& \varepsilon_{\mathrm{f}} + \varepsilon_{\mathrm{l}} u_{IJ} h^{IJ} = \varepsilon_{\mathrm{tot}} \; , \\
		\mathcal{N} &=& n_{\mathrm{f}} + n_{\mathrm{l}} u_{IJ} h^{IJ} = n_{\mathrm{tot}} \; ,
	\end{eqnarray}
where we have again dropped terms that will not be relevant in the linearised limit. The vanishing of corrections to $\mathcal{E}$, $\mathcal{P}^{\mu}$ and $\mathcal{N}$ is equivalent to working in the Landau frame which is defined by
	\begin{eqnarray}
		u^{\mu} T_{\mu \nu}  = - \varepsilon_{\mathrm{tot}} u_{\nu} \; , \qquad J^{\mu} u_{\mu} = -n_{\mathrm{tot}} \; .
	\end{eqnarray}
In this frame, our equilibrium constitutive relations take the form
\begin{subequations}
	\begin{eqnarray}
    	   \mathcal{E} &=& \left(\varepsilon_{\mathrm{f}} + (u_{IJ} h^{IJ}) \varepsilon_{\mathrm{l}} \right)  \; , \qquad \mathcal{N} = \left( n_{\mathrm{l}} + n_{\mathrm{l}} ( u_{IJ} h^{IJ} ) \right) \; ,  \qquad \\
	   \mathcal{P} &=& \left( P_{\mathrm{f}} + ( u_{IJ} h^{IJ} ) P_{\mathrm{l}} \right) - B \left( m_{\mathrm{f}} +  ( u_{IJ} h^{IJ} ) m_{\mathrm{l}} \right)  - \frac{r_{IJ}}{2} e^{I}_{\alpha} \Pi^{\alpha \beta} e^{J}_{\beta} - \tilde{\chi}_{\Omega} B \Omega \; , \qquad \\
	 \mathcal{P}^{\mu} &=& 0 \; , \qquad \mathcal{T}^{\mu \nu} = - r_{IJ} e^{I}_{\alpha} \Pi^{\alpha \langle \mu}  \Pi^{\nu \rangle \beta} e^{J}_{\beta}  \; , \\
	\mathcal{J}^{\mu} &=& \epsilon^{\mu \nu \rho} u_{\nu} \partial_{\rho} \left( m_{\mathrm{f}} + ( u_{IJ} h^{IJ} ) m_{\mathrm{l}} \right)  - \left( m_{\mathrm{f}} + B \tilde{\chi}_{T} \right) \epsilon^{\mu \nu \rho} u_{\nu} \frac{\partial_{\rho} T}{T}  \nonumber \\
				   &\;& + B \tilde{\chi}_{E}  \epsilon^{\mu \nu \rho} u_{\nu} E_{\rho} \; , 
	\end{eqnarray}
	\end{subequations}
where again we have dropped terms that will not be relevant for the linearised constitutive relations. The terms $\tilde{\chi}_{T}$, $\tilde{\chi}_{E}$ and $\tilde{\chi}_{\Omega}$ can be expressed entirely in terms of thermodynamic quantities (in particular $g_{1}$, $g_{2}$ and $g_{3}$). While the structure of our equilibrium constitutive relations bears more than a passing resemblance to those given in Ref. \cite{Jensen:2011xb} the reader should remember that our fluid is not parity violating and the magnetic field is $\mathcal{O}(\partial^{0})$. Our expressions also compare favourably with those in Ref. \cite{Armas:2020bmo} on setting $B=0$ up to a minor caveat; in Ref. \cite{Armas:2020bmo} the published expression for the stress tensor is not in the Landau frame.}

{\ With the Landau frame equilibrium expressions to hand we can add back terms proportional to the equilibrium constraints \eqref{Eq:B=0Killing} and \eqref{Eq:Killing}. The difference from zero of these constraints is always taken to be at least $\mathcal{O}(\partial)$. For example, $\Pi^{\mu \nu} \partial_{\nu} \phi^{I} \sim \mathcal{O}(\partial^{0})$ while $u^{\mu} \partial_{\mu} \phi^{I} \sim \mathcal{O}(\partial)$. Hence we find
\begin{subequations}
	\begin{eqnarray}
		\label{Eq:TJdissipative}
		\mathcal{T}^{\mu \nu}_{\mathrm{non}.} &=& - 2 \eta \sigma^{\mu \nu}  - 2 \tilde{\eta} \tilde{\sigma}^{\mu \nu} \; , \qquad \mathcal{P}_{\mathrm{non}.} = - \zeta \theta  \; , \\
		J^{\mu}_{\mathrm{non}.} &=& \sigma_{n}^{\mu \nu} \left( E_{\nu} - T \Pi\indices{_{\nu}^{\rho}} \partial_{\rho} \left( \frac{\mu}{T} \right) \right)
				    + \gamma_{1}^{IJ} P^{\mu}_{I} u^{\nu} e_{\nu J}  + \tilde{\gamma}_{1}^{IJ} P^{\mu}_{I} u^{\nu} \partial_{\nu} \psi_{J} \; , 
	\end{eqnarray}
\end{subequations}
with
	\begin{eqnarray}
		\tilde{\sigma}^{\mu \nu} &=& B \left( \epsilon^{\mu \rho \sigma} u_{\sigma} \sigma\indices{_{\rho}^{\nu}} + \epsilon^{\nu \rho \sigma} u_{\sigma} \sigma\indices{_{\rho}^{\mu}} \right) \; , 
	\end{eqnarray}
while 
	\begin{eqnarray}
		\label{Eq:Goldstonedissipative}
		K_{I}^{\mathrm{non}.} &=& \gamma'_{I J} P^{J \mu} \left( E_{\mu} - T \Pi\indices{_{\mu}^{\nu}} \partial_{\nu} \left( \frac{\mu}{T} \right) \right) - \sigma^{'\Phi}_{IJ}  u^{\mu} e_{\mu}^{J} + \sigma^{\times}_{IJ} u^{\nu} \partial_{\nu} \psi^{J} \; , \\
		L_{I}^{\mathrm{non}.} &=& \tilde{\gamma}'_{I J} P^{J \mu} \left( E_{\mu} - T \Pi\indices{_{\mu}^{\nu}} \partial_{\nu} \left( \frac{\mu}{T} \right) \right) -  \sigma^{'\times}_{IJ} u^{\mu} e_{\mu}^{J} + \sigma^{\Phi}_{IJ} u^{\nu} \partial_{\nu} \psi^{J}  \; .	
	\end{eqnarray}
Following \cite{Amoretti:2020mkp} each of our transport coefficients contains a longitudinal and a Hall piece e.g.
	\begin{eqnarray}
		\sigma^{IJ} &=& \sigma_{(\mathrm{L})} \delta^{IJ} +  \sigma_{(\mathrm{H})} F^{IJ} \; , \qquad F^{IJ} = F^{\mu \nu} e_{\mu}^{I} e_{\nu}^{J} \; . 
	\end{eqnarray}
This decomposition with the explicit field strength tensor is appropriate for a spatial parity invariant theory. In particular, fluids whose microscopic theory breaks spatial parity invariance contain Hall terms that are proportional to $\epsilon^{\mu \nu \rho} u_{\rho}$ and not $B \epsilon^{\mu \nu \rho} u_{\rho} \sim F^{\mu \nu}$. Positivity of entropy production \eqref{Eq:entropypositivity} then imposes constraints on the longitudinal transport coefficients. For example;
	\begin{eqnarray}
		\eta, \; \zeta \geq 0 \; . 
	\end{eqnarray}
As these constraints are somewhat complex in our case, and will not play a role in our story, we do not record them here suffice to say that longitudinal components are constrained while Hall components are not.}

\section{The hydrodynamic Green's functions}
\label{sec:Green}

% What are the Ward identities and their importance
{\noindent The response of a system to turning on a perturbing operator about some particular ground state is encoded in the Green's function. While the precise form of the Green's function is dependent on both the system and operator they are nonetheless constrained by any discrete or continuous symmetry of the problem. In particular,  current operators associated with continuous symmetries such as the stress tensor or $U(1)$ charge current must obey the ``Ward identities". We shall now obtain the linearised constitutive relations for out-of-equilibrium fluids and by imposing the Ward identities constrain transport coefficients.}

\subsection{Ward identities}

{\noindent We suppose that there exists some microscopic action $S[\phi;g,A]$ describing our system where $g$ is the metric, $A$ an external gauge field, $K_{\mathrm{ext.}}$ an external source for the effective translation breaking scalars $\phi^{I}$ and $\varphi$ denotes all microscopic matter fields including those that break translation invariance. Consequently we can define the generating functional for $n$-point functions of the stress tensor, $U(1)$ charge current and pseudo-Goldstone field to be
	\begin{eqnarray}
		\label{Eq:GeneratingFunctional}
		e^{W[\varphi;g,A,K_{\mathrm{ext}.}]} = Z[\varphi;g,A,K_{\mathrm{ext}.}] = \int D\varphi \; e^{i S[\varphi;g,A] + \int d^{2+1}x \; \sqrt{-g} K^{\mathrm{ext}.}_{I} \phi^{I}} \; . 
	\end{eqnarray}
Derivatives of this expression with suitable boundary conditions yield the various time ordered correlators. In particular, the SEM tensor, $U(1)$ charge current and scalar operator one-point functions are given by
	\begin{align}
		& J^{\mu} = \langle J^{\mu} \rangle = \left. \frac{\delta W}{\delta A_{\mu}} \right|_{(g,A,K_{\mathrm{ext}.})=(\eta,A_{0},0)} \; , \; \; T^{\mu \nu} = \langle T^{\mu \nu} \rangle = \left. \frac{\delta W}{\delta g_{\mu \nu}} \right|_{(g,A,K_{\mathrm{ext}.})=(\eta,A_{0},0)} \; , \; \; \nonumber \\
		& \phi^{I} = \langle \phi^{I} \rangle = \left. \frac{\delta W}{\delta K^{\mathrm{ext}.}_{I}} \right|_{(g,A,K_{\mathrm{ext}.})=(\eta,A_{0},0)} \qquad \qquad
	\end{align}
where $\eta$ is the Minkowski metric and $A_{0}^{\mu}$ some background value for the gauge field. The hydrodynamic equations of motion subsequently follow from diffeomorphism and gauge invariance of the generating functional \eqref{Eq:GeneratingFunctional}.}

{\ We also have cause to consider the various two point functions given by a second variation of the generating functional with respect to the metric, external gauge field and translation breaking scalar source. An in depth discussion of how to obtain these Ward identities can be found in Ref. \cite{Herzog:2009xv}. For our purposes however it is sufficient to assume homogeneity and following a convention where
   \begin{eqnarray}
     f(t,\vec{x}) = \int \frac{d^{2}k d\omega}{(2 \pi)^3} f(\omega,\vec{k}) e^{- i ( \omega t - i \vec{k} \cdot \vec{x} )} \; ,
    \end{eqnarray}
one finds at zero-wavevector ($\vec{k}=0$) the following relations
    \begin{subequations}
        \label{Eq:2ptWard}
        \begin{eqnarray}
            \label{Eq:ExplicitWardIdentity1}
            i \omega \langle Q^{i} Q^{j} \rangle &=& - \left( i \omega \mu \delta\indices{^{i}_{k}} - F\indices{^{i}_{k}} \right) \langle Q^{k} J^{j} \rangle + m \langle Q^{i} \phi^{J} \rangle \delta^{jJ}
            - i \omega \left( \chi_{\pi \pi} - \mu n \right) \delta^{ij} \; , \qquad \\
            \label{Eq:ExplicitWardIdentity2}
            i \omega \langle Q^{i} J^{j} \rangle &=& - \left( i \omega \mu \delta\indices{^{i}_{k}} - F\indices{^{i}_{k}} \right) \langle J^{k} J^{j} \rangle + m \langle J^{i} \phi^{J} \rangle \delta^{jJ} - i \omega n \delta^{ij} \; , \\
            \label{Eq:ExplicitWardIdentity3}
            i \omega \langle Q^{i} \phi^{J} \rangle &=& - \left( i \omega \mu \delta\indices{^{i}_{k}} - F\indices{^{i}_{k}} \right) \langle J^{k} \phi^{J} \rangle - m \langle \phi^{I} \phi^{J} \rangle \delta^{iI} + \delta^{iJ} \; ,
        \end{eqnarray}
    \end{subequations}
where $Q^{i} = T^{it} - \mu J^{i}$ is the canonical heat current and $m$ is the pseudo-Goldstone mass. We use the canonical heat rather than the entropic heat (that which enters the entropy production equation) as in principle one needs to work at all orders in derivatives to determine the one-point Ward identity satisfied by the entropic heat current, and subsequently the two-point Ward identity.}

{\ We note the existence of a ladder structure in \eqref{Eq:2ptWard} which reduces the number of independent correlators from six to three: $\langle \phi^{I} \phi^{J} \rangle$, $\langle J^{i} \phi^{J} \rangle$ and $\langle J^{i} J^{j} \rangle$ i.e. every other correlator can be determined in terms of these three. As a consequence one finds that the leading terms in the $\omega \to 0$ limit of the original six correlators are all contained in the low frequency expansion of the independent correlators. Hence, knowing the DC values of all the correlators is as good as knowing the low frequency expansion of the independent correlators. Comparing the hydrodynamic expressions at low frequencies with what is imposed by the Ward identities \eqref{Eq:2ptWard} allows us to fix the hydrodynamic transport coefficients analytically when we know the DC terms analytically.}

\subsection{Onsager relations}

{\noindent Having determined the two-point functions, one can also impose the so-called Onsager relations to further constrain the dissipative transport coefficients. These discrete symmetry relations tie together particular transport coefficients to each other by asking how the two-point functions transform under time reversal invariance. Keeping in mind that $B$ is a time reversal odd scalar\cite{Kovtun:2012rj} one finds that
	\begin{subequations}
	\label{Eq:OnsagerRelations}
	\begin{eqnarray}
		\sigma^{'\Phi}_{IJ} &=& \frac{w_{\mathrm{f}}}{\chi_{\pi \pi}} \left( \sigma^{\times}_{IJ} - \sigma^{\phi}_{IJ} \right) - \frac{P_{\mathrm{l}}^{2}}{w_{\mathrm{f}} \chi_{\pi \pi}} \left( \sigma^{\times}_{IJ} + \sigma^{\Phi}_{IJ} \right)  \; , \\
		\sigma^{'\times}_{IJ} &=& \frac{\chi_{\pi \pi}}{w_{\mathrm{f}}} \left( \sigma^{\times}_{IJ} + \sigma^{\Phi}_{IJ} \right) \; , \qquad \tilde{\gamma}'_{IJ} = -  \tilde{\gamma}_{IJ} \; , \qquad \\
		\gamma'_{IJ} &=& \frac{w_{\mathrm{f}}}{\chi_{\pi \pi}} \gamma_{IJ} + \tilde{\gamma}_{IJ} + \frac{P_{\mathrm{l}} q_{\mathrm{f}}}{\chi_{\pi \pi}} \delta_{IJ} \; , 
	\end{eqnarray}
	\end{subequations}
where we have defined
	\begin{subequations}
	\begin{eqnarray}
		\chi_{\pi \pi} &=&  \left( \varepsilon_{\mathrm{f}} + P_{\mathrm{f}} + P_{\mathrm{l}} - 2 \mathfrak{m}_{\mathrm{f}} B^2 \right) \; , \\
		w_{\mathrm{f}} &=& \chi_{\pi \pi} - P_{\mathrm{l}} \; ,
	\end{eqnarray}
	\end{subequations}
to be the momentum susceptibility and free enthalpy respectively while $\sigma^{\phi}_{IJ}$ is the crystal diffusivity\cite{Armas:2020bmo}. Given the above relations, upon setting $P_{\mathrm{l}}=0$ and $B=0$ we find that the our constitutive relations become a natural generalisation to the relativistic case of the expressions given in Ref. \cite{Armas:2021vku}.}

\subsection{Diffusion and linearised, out-of-equilibrium constitutive relations}

\subsubsection{With only a magnetic field}

    \begin{figure}[t]
 \centering
  \begin{subfigure}
  \centering
  \includegraphics[width=0.43\textwidth]{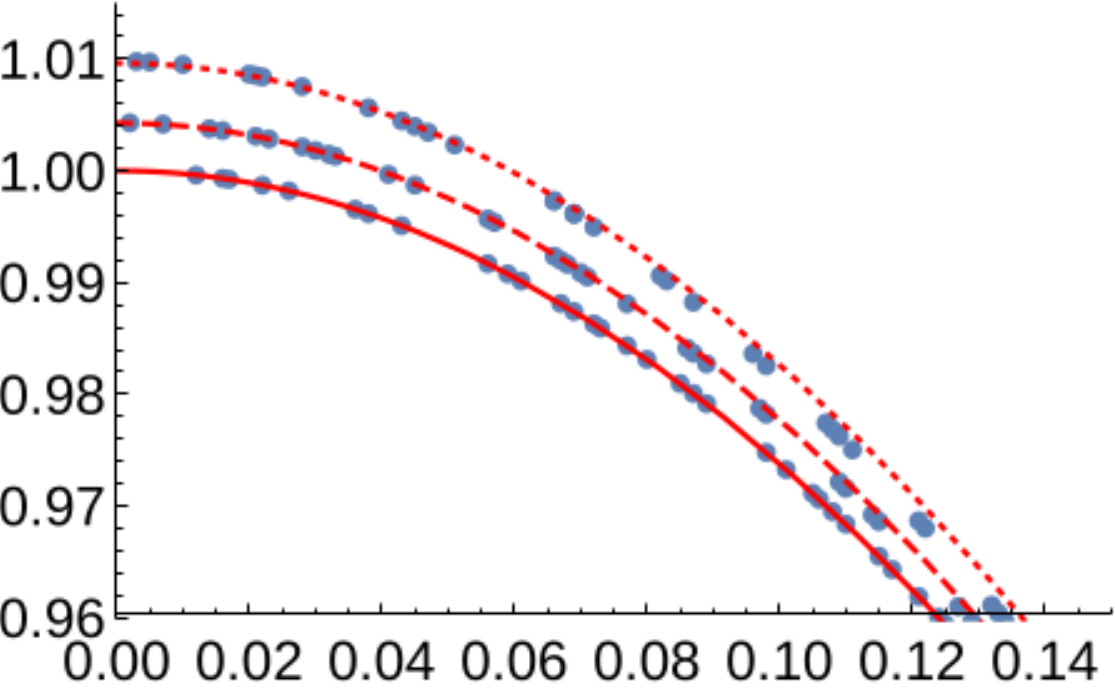}
 \end{subfigure} \qquad 
 \begin{subfigure}
  \centering
  \includegraphics[width=0.43\textwidth]{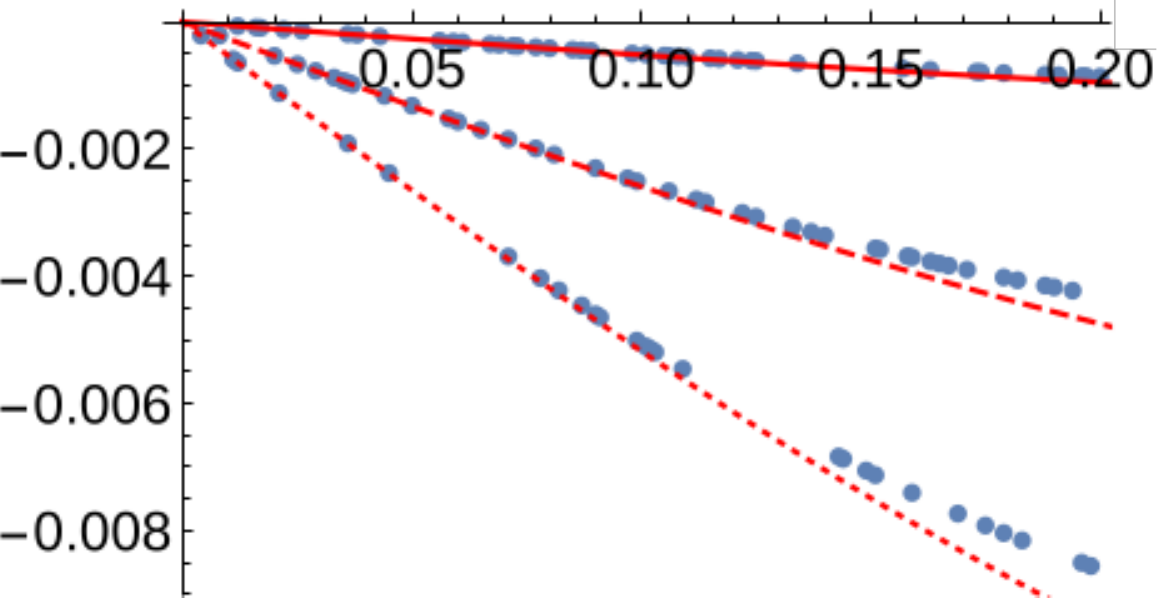}
 \end{subfigure} \qquad
 \begin{picture}(100,0)
  \put(-110,121){\small{$\mathrm{Im}[\sigma_{+}^{\mathrm{inc}}]$}}
  \put(38,30){\small{$\frac{\pi n_{\mathrm{f}}}{s}$}}
  \put(75,105){\small{$\mathrm{Re}[\sigma_{+}^{\mathrm{inc}}]$}}
  \put(220,95){\small{$\frac{\pi n_{\mathrm{f}}}{s}$}}
 \end{picture}
 \vskip-1em
    \caption{Plots of the constant term in the Laurent expansion of the charge correlator about the hydrodynamic pole, denoted $\sigma_{+}^{\mathrm{inc.}}$ against the charge density. The blue dots are data and the red lines analytic expressions obtained from hydrodynamics. \textbf{Left:} The imaginary part of $\sigma_{+}^{\mathrm{inc.}}$ against our analytic expression for $\sigma_{0}$. The three red lines represent $\pi B/s=1/1000$ (solid), $1/25$ (dashed) and $3/50$ (dotted). \textbf{Right:} The real part of $\sigma_{+}^{\mathrm{inc.}}$ against our analytic expression for $\tilde{\sigma}_{\mathrm{H}}$. The three red lines represent $\pi B/s=1/1000$ (solid), $5/1000$ (dashed) and $10/1000$ (dotted).}
    \label{Fig:Incoherentconductivity}
\end{figure}
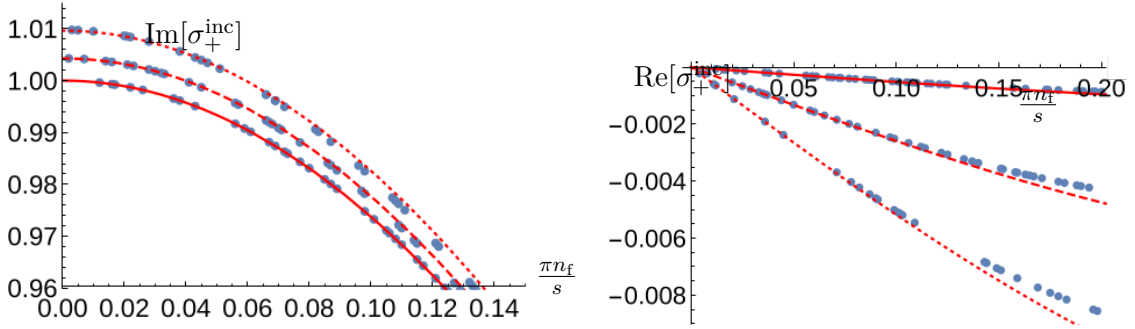

{\noindent The simplest holographic model for hydrodynamics in $(2+1)$-dimensions with an external magnetic field is the dyonic black hole\cite{Hartnoll:2007ai,Hartnoll:2007ih,Hartnoll:2007ip,Amoretti:2020mkp}. Early studies\cite{Hartnoll:2007ip} initially missed important Hall transport coefficients in their constitutive relations. It turns out such a coefficient is a necessary consequence of the Ward identities (as was demonstrated recently\cite{Amoretti:2020mkp}) in the presence of a non-zero magnetic field in global thermodynamic equilibrium. This was the cause of an infamous discrepancy between holographic and hydrodynamic calculations where the latter could only reproduce the DC thermal conductivities at leading order\cite{Hartnoll:2007ih} in $(n/B)^2$.}

{\ The electric current-current correlator for these systems can be obtained by linearising our hydrodynamic expressions with all the translation breaking scalar terms set to zero. As the electric conductivity, $\sigma_{ij}(\omega)$, is the only independent two point function at zero wavevector we only need to specify its form:
	\begin{subequations}
	\begin{eqnarray}
		\label{Eq:RetardedDyonic}
		\hat{\sigma}(\omega) &=& \left(  \sigma_{n,(\mathrm{L})} \hat{\mathbbm{1}}_{2} + \tilde{\sigma}_{n,(\mathrm{H})} \hat{F} \right) - \hat{\Lambda}^{-1} \left( n \hat{\mathbbm{1}}_{2} + \hat{F} \cdot \left( \sigma_{n,(\mathrm{L})} \mathbbm{1}_{2} + \tilde{\sigma}_{n,(\mathrm{H})}\hat{F} \right) \right)^2 \; , \\
		\label{Eq:dyonicQNMs}
		\hat{\Lambda} &=& i \omega \left( \varepsilon_{\mathrm{f}} + P_{\mathrm{f}} - 2 \mathfrak{m}_{\mathrm{f}} B^2 \right) \hat{\mathbbm{1}}_{2}  + \hat{F} \cdot \left( n \hat{\mathbbm{1}}_{2} + \hat{F} \cdot \left( \sigma_{n,(\mathrm{L})} \mathbbm{1}_{2} + \tilde{\sigma}_{n,(\mathrm{H})} \hat{F} \right) \right) , \qquad \\
		\label{Eq:effectivehallconductivity}
		\tilde{\sigma}_{n,(\mathrm{H})} &=& \sigma_{n,(\mathrm{H})} - \tilde{\chi}_{E} \; , 
	\end{eqnarray}
	\end{subequations}
where the superscript $\hat{}$ indicates a $(2 \times 2)$-matrix quantity.}

{\ We notice the appearance of $\tilde{\chi}_{E}$ given in \eqref{Eq:effectivehallconductivity}. Such a term is missing in even the latest treatments\cite{Amoretti:2020mkp} and as far as the authors are aware its value has not been obtained for the dyonic black hole in the holographic literature. It enters into the current-current correlator in precisely the same manner as $\sigma_{n,(\mathrm{H})}$ and thus the quantity identified as the incoherent Hall conductivity in Ref. \cite{Amoretti:2020mkp} is \eqref{Eq:effectivehallconductivity}. To isolate the effects of $\tilde{\chi}_{E}$ it is necessary to look at other correlators that are given by a variation with respect to $\delta \mu$, rather than $\delta E_{x}$, such as $\langle J^{i} \rho \rangle$.}

{\ Given \eqref{Eq:RetardedDyonic}, the transport coefficients $\sigma_{n,(\mathrm{L})}$ and $\tilde{\sigma}_{n,(\mathrm{H})}$ can be obtained by comparing the low frequency expansion of \eqref{Eq:RetardedDyonic} against the expansion dictated by the Ward identities \eqref{Eq:2ptWard}. This fixes  $\sigma_{n,(\mathrm{L})}$ and $\tilde{\sigma}_{n,(\mathrm{H})}$ uniquely in terms of the longitudinal and Hall DC thermal conductivities. That $\tilde{\sigma}_{n,(\mathrm{H})}$ must be non-zero can be demonstrated by extracting the constant term of the Laurent expansion about the hydrodynamic pole ($\sigma_{+}^{\mathrm{inc}}$) which at small $B$ has the form
	\begin{eqnarray}
		\label{Eq:ExpansionConductivity}
		\sigma_{+}^{\mathrm{inc}} = i \left[ \sigma_{n,(\mathrm{L})} \right]_{B=0} + B \left[ \tilde{\sigma}_{n,(\mathrm{H})}  \right]_{B=0}  + \mathcal{O}(B^2) \; .
	\end{eqnarray}
This quantity is displayed in fig. \ref{Fig:Incoherentconductivity} where, importantly, we see that the real part is non-zero and therefore $\tilde{\sigma}_{n,(\mathrm{H})} \neq 0$.}

\subsubsection{With translation breaking scalars}

{\noindent Turning now to the case of broken translation invariance. We linearise our hydrodynamic expressions - having imposed the Onsager relations \eqref{Eq:OnsagerRelations} - around a zero velocity background where $T$, $\mu$, $B$ are constants and 
	\begin{eqnarray}
		\phi^{i} &=& x^{i} - \delta \phi^{i} \; , \qquad \Phi^{i} = x^{i} \;, \qquad \psi^{i} = - \delta \phi^{i} \; . 	
	\end{eqnarray}
We Laplace and Fourier transform then take the zero wavevector limit. Fluctuations of the chemical potential and temperature, $\delta \mu$ and $\delta T$, will play no role in our current analysis. In what follows we take $\hat{\sigma}_{\Phi}=\hat{\sigma}_{\times}=\hat{\tilde{\gamma}} = 0$. The reason for doing this is that we compare against a holographic model and to date the values of these coefficients are not known. Nevertheless we do know that they must be small as the holographic result without them matches well the hydrodynamic result within the regime where hydrodynamics is expected to apply.}

{\ Following Ref. \cite{Amoretti:2021lll}, let us define the AC transport coefficients matrices
    \begin{eqnarray}
	\label{Eq:DefinitionsofDC1}
	\left(\sigma^{ij}, \alpha^{ij}, \varpi^{iJ}\right)(\omega) &=& \frac{1}{i\omega}\left(  \langle J^{i} J^{j} \rangle , \langle Q^{i} J^{j} \rangle,  \langle J^{i} \phi^{J} \rangle \right) \; , \\
	\label{Eq:DefinitionsofDC2}
	\left(\kappa^{ij}, \zeta^{IJ}, \theta^{iJ}\right)(\omega) &=& \frac{1}{i\omega}\left(  \langle Q^{i} Q^{j} \rangle , \langle \phi^{I} \phi^{J} \rangle - \frac{1}{m} \delta^{IJ},  \langle Q^{i} \phi^{J} \rangle \right) \; ,
\end{eqnarray}
which can be expanded into longitudinal and Hall components,
\begin{eqnarray}
	\label{Eq:Thermalconductivities}
	(\hat\sigma_{n},\hat\alpha,\hat\kappa,\hat\varpi,\hat\zeta,\hat\theta)(\omega)=(\sigma_{n},\alpha,\kappa,\varpi,\zeta,\theta)_{(\mathrm{L})}(\omega)\mathbbm{1}_{2}+(\sigma_{n},\alpha,\kappa,\varpi,\zeta,\theta)_{(\mathrm{H})}(\omega) \hat{F} \; . 
\end{eqnarray}
With these definitions to hand we find the three independent AC correlators are then given by
\begin{subequations}
	\label{Eq:ExplicitACconductivities}
	\begin{align}
		\hat\sigma_{n}(\omega)&=\hat\Xi^{-1}\cdot\left[\omega_{0}^2 \chi_{\pi \pi} \omega w_f\hat\sigma_{n}+\omega n_{\mathrm{f}}^2\hat\sigma_\phi-i(\omega^2-\omega_0^2)\chi_{\pi\pi}\hat\sigma'-n_{\mathrm{f}}\left(i\omega_0^2 \chi_{\pi \pi}\hat\rho+\omega \hat{F} \cdot\hat\sigma'\right)\right]\;,\\
		\hat\varpi(\omega)&=\hat\Xi^{-1}\cdot\left(\omega w_f\hat\gamma+i(n_{\mathrm{f}}\hat\sigma_\phi-\hat{F}\cdot\hat\sigma')\right)\;,\\
		\hat\zeta(\omega)&=\frac{1}{\omega_0^2\chi_{\pi\pi}}\hat\Xi^{-1}\cdot\left(\omega\chi_{\pi\pi}\hat\sigma_\phi-\omega P_l\left(\hat{F}\cdot\hat\rho+i\omega w_f\mathbbm{1}_{2}\right)+i\hat{F}\cdot(n_{\mathrm{f}}\hat\sigma_\phi-\hat{F}\cdot\hat\sigma')\right) \; ,
	\end{align}
\end{subequations}
where we have defined
\begin{subequations}
	\begin{align}
	\hat\sigma'&=\hat\gamma^2+\hat\sigma_{n}\cdot\hat\sigma_\phi \; , \qquad \hat\rho=2\hat\gamma+\hat{F}\cdot\hat\sigma_{n}-n_f\mathbbm{1}_2 \; , \\
		\hat\Xi&=\omega_{0}^2 \chi_{\pi \pi} \left(\omega w_f\mathbbm{1}_{2}-i \hat{F}\cdot\hat\rho\right)+\omega n_{\mathrm{f}}\hat{F}\cdot\hat\sigma_\phi-i(\omega^2-\omega_0^2)\chi_{\pi\pi}\hat\sigma_\phi-\omega \hat{F}^2\cdot\hat\sigma'\;. \label{denominatore}
	\end{align}
\end{subequations}
Consequently, by employing the Ward identities \eqref{Eq:2ptWard}  the three hydrodynamic transport matrices $\hat\sigma_{n}$, $\hat\sigma_{\phi}$ and $\hat\gamma$ are expressed in terms of the DC values of the electric, thermoelectric and thermal conductivity $\hat\sigma_{n}(0)$, $\hat\alpha(0)$ and $\hat\kappa(0)$ hydrodynamic transport coefficients: 
\begin{subequations}
	\label{Eq:ExplicitACtransportcoeffs}
	\begin{align}
		\hat\sigma_{n}&=-\hat\Psi^{-1}\cdot\hat\pi(0)\;,\\
		\hat\sigma_\phi&=\hat\Psi^{-1}\cdot\left[w_f^2\mathbbm{1}_{2}
		+\left(\hat{F}\cdot\hat\pi(0)-2w_f(\hat\alpha(0)+\mu\hat\sigma_{n}(0))\right)\cdot \hat{F}\right]+n_{\mathrm{f}}\hat{F}\;,\\
		\hat\gamma&=\hat\Psi^{-1}\cdot\left[\hat{F}\cdot\hat\pi(0)-w_f(\hat\alpha(0)+\mu\hat\sigma_{n}(0))\right]+n_{\mathrm{f}}\mathbbm{1}_{2}\;,\\
		\hat\Psi&=\mu^2\hat\sigma_{n}(0)+2\mu\hat\alpha(0)+\hat\kappa(0)\;,
	\end{align}
\end{subequations}
where we have defined
\begin{equation}
	\hat\pi(0)=\hat\alpha^2(0)-\hat\kappa(0)\cdot\hat\sigma(0)\; .
\end{equation}
In Ref. \cite{Amoretti:2021lll} these holographic correlators have been compared to a holographic model in which the breaking of translations is realised either spontaneously or pseudo-spontaneously by means of scalar operators which acquire a non-trivial position dependent profile. The agreement between the hydrodynamic and the holographic results holds up to four significant digits.}

\section{Future work}

{\noindent In this review we have provided a (naturally non-exhaustive) summary of the literature on charged hydrodynamics in $(2+1)$-dimensions with an external magnetic field, and a particular mechanism for broken translation invariance. We have highlighted some of the interesting research that has been done and noted issues with the literature; nevertheless, these issues bare repeating.}

{\ As yet we lack a holographic model where the additional transport coefficients given in \eqref{Eq:TJdissipative} and \eqref{Eq:Goldstonedissipative} have been determined. Certain of these transport coefficients and thermodynamic parameters are obtainable using Kubo formulae - however, when this is not the case it is dangerous to extract the unknown coefficients using a quasinormal mode analysis, as is sometimes done in the literature. Hence why we have championed the Ward identity approach whenever it can be applied. The reason for this  caution is that it is difficult to separate contributions to the quasinormal modes coming from the additional coefficients displayed in \eqref{Eq:TJdissipative} and \eqref{Eq:Goldstonedissipative}, compared to higher derivative corrections. Even more so as the number of transport coefficients multiplies.}

{\ One of the key features missing from previous work is a complete treatment of electric and magnetic fields; especially if the fluid has a non-zero velocity in global thermodynamic equilibrium. In particular it is very natural from the experimental point of view to have a steady flow of charge in the presence of a constant, external, electric field - such is how most electronic devices work when attached to a battery. Naturally, potential devices that can be described by a fluid will have broken boost invariance and one can expect the spatial velocity to be an important piece of the definition of equilibrium for such models.}

{\ Additionally, the literature is too reliant upon models where the microscopic mechanism of momentum loss is known. This should be compared to the robustness of the Drude model where it is not necessary to know the precise mechanism responsible for momentum loss to derive useful results. It would be quite interesting to understand the constraints one may impose on generic models with momentum relaxation when one deliberately chooses to remain agnostic about the microscopic mechanism. A step in this direction will be the subject of upcoming work\cite{electrohydro}.}

\appendix

\section*{Acknowledgments}

AA  has been partially supported by the ``Curiosity Driven Grant 2020'' of the University of Genoa and the INFN Scientific Initiative SFT: ``Statistical Field Theory, Low-Dimensional Systems, Integrable Models and Applications''. This project has also received funding from the European Union's Horizon $2020$ research and innovation programme under the Marie Sk\l{}odowska-Curie grant agreement No. $101030915$. DB would like to acknowledge discussions with Blaise Gouteraux, Luca Martinoia, Ioannis Matthaiakakis, Niko Jokela, Ashish Shukla, Benjamin Withers and Vaios Ziogas on the issues presented in this work.

\bibliographystyle{unsrt} 
\bibliography{bib1}

\end{document}